\documentclass[apj]{emulateapj}
\usepackage{color}

\shorttitle{Broad-Lined Supernova 2016coi with Helium Envelope} \shortauthors{Yamanaka et al.}

\begin{document}

\title{Broad-Lined Supernova 2016coi with Helium Envelope}

\author{Masayuki \textsc{Yamanaka}\altaffilmark{1},    %1 Konan
         Tatsuya \textsc{Nakaoka}\altaffilmark{2},     %2 Hiroshima
         Masaomi \textsc{Tanaka}\altaffilmark{3},      %3 NAOJ,mitaka
         Keiichi \textsc{Maeda}\altaffilmark{4,5},     %4 Kyoto, %5IPMU
         Satoshi \textsc{Honda}\altaffilmark{6},       %6 NHAO
         Hidekazu \textsc{Hanayama}\altaffilmark{7},   %7 IAO
	 Tomoki \textsc{Morokuma}\altaffilmark{8},     %8 Tokyo Mitaka
         Masataka \textsc{Imai}\altaffilmark{9},       %9 Hokudai1
         Kenzo \textsc{Kinugasa}\altaffilmark{10},     %10 NAOJ,nobeyama
         Katsuhiro L. \textsc{Murata}\altaffilmark{11},%11 Nagoya 
         Takefumi \textsc{Nishimori}\altaffilmark{12}, %12 Kagoshima 
         Osamu \textsc{Hashimoto}\altaffilmark{13},    %13Gunma
         Hirotaka \textsc{Gima}\altaffilmark{12},
         Kensuke \textsc{Hosoya}\altaffilmark{6},
         Ayano \textsc{Ito}\altaffilmark{12},
         Mayu \textsc{Karita}\altaffilmark{6},
         Miho \textsc{Kawabata}\altaffilmark{2},           
         Kumiko \textsc{Morihana}\altaffilmark{6},
         Yuto \textsc{Morikawa}\altaffilmark{12},
         Kotone \textsc{Murakami}\altaffilmark{12}, 
         Takahiro \textsc{Nagayama}\altaffilmark{12}, 
         Tatsuharu \textsc{Ono}\altaffilmark{14},      %14 Hokudai2
         Hiroki \textsc{Onozato}\altaffilmark{15},     %15 Tohoku
         Yuki \textsc{Sarugaku}\altaffilmark{16},      %16 Kiso
         Mitsuteru \textsc{Sato}\altaffilmark{17},     %17 Hokudai3
         Daisuke \textsc{Suzuki}\altaffilmark{18},     %18 NASA
         Jun \textsc{Takahashi}\altaffilmark{6}, 
         Masaki \textsc{Takayama}\altaffilmark{6}, 
         Hijiri \textsc{Yaguchi}\altaffilmark{6},
         Hiroshi \textsc{Akitaya}\altaffilmark{2,19},  %9 Hiroshima, %19 center
         Yuichiro \textsc{Asakura}\altaffilmark{20},   %20 Nagoya2
         Koji S. \textsc{Kawabata}\altaffilmark{2,19},
         Daisuke \textsc{Kuroda}\altaffilmark{21},     %21 OAO 
         Daisaku \textsc{Nogami}\altaffilmark{4},
 	 Yumiko \textsc{Oasa}\altaffilmark{22},        %22 Saitama
         Toshihiro \textsc{Omodaka}\altaffilmark{12}, 
         Yoshihiko \textsc{Saito}\altaffilmark{23},    %23 Tokodai
         Kazuhiro \textsc{Sekiguchi}\altaffilmark{3}, 
         Nozomu \textsc{Tominaga}\altaffilmark{1,5}, 
         Makoto \textsc{Uemura}\altaffilmark{2,19}, and   
         Makoto \textsc{Watanabe}\altaffilmark{24}.  %23 Okayama Rikadai
%        Michitoshi \textsc{Yoshida}\altaffilmark{3,9},
}

\altaffiltext{1}{Department of Physics, Faculty of Science and Engineering, 
Konan University, Okamoto, Kobe, Hyogo 658-8501, Japan; yamanaka@center.konan-u.ac.jp} 
\altaffiltext{2}{Department of Physical Science, Hiroshima University, Kagamiyama 1-3-1, Higashi-Hiroshima 739-8526, Japan} 
\altaffiltext{3}{National Astronomical Observatory of Japan, National Institutes of Natural Sciences, 
Osawa, Mitaka, Tokyo 181-8588, Japan}
\altaffiltext{4}{Department of Astronomy, Graduate School of Science, Kyoto University, Sakyo-ku, Kyoto 606-8502, Japan}
\altaffiltext{5}{Kavli Institute for the Physics and Mathematics of the Universe (WPI), 
The University of Tokyo, 5-1-5 Kashiwanoha, Kashiwa, Chiba
277-8583, Japan}
\altaffiltext{6}{Nishi-Harima Astronomical Observatory, Center for Astronomy, University of Hyogo, 407-2 Nishigaichi, Sayo-cho, Sayo, Hyogo 679-5313, Japan}
\altaffiltext{7}{Ishigakijima Astronomical Observatory, National Astronomical Observatory of Japan, National Institutes of Natural Sciences, 1024-1 Arakawa, Ishigaki, Okinawa 907-0024, Japan}
\altaffiltext{8}{Institute of Astronomy, Graduate School of Science, The University of Tokyo, 2-21-1 Osawa, Mitaka, Tokyo 181-0015, Japan}
\altaffiltext{9}{Department of Cosmosciences, Graduate School of Science, Hokkaido University, Kita 10 Nishi8, Kita-ku, Sapporo 060-0810, Japan}
\altaffiltext{10}{Nobeyama Radio Observatory, National Astronomical Observatory of Japan, National Institutes of Natural 
Sciences, 462-2 Nobeyama, 
Minamimaki, Minamisaku, Nagano 384-1305, Japan}
\altaffiltext{11}{Department of Astrophysics, Nagoya University, Chikusa-ku, Nagoya 464-8602, Japan}
\altaffiltext{12}{Graduate School of Science and Engineering, Kagoshima University, 1-21-35 Korimoto, Kagoshima 890-0065, Japan}
\altaffiltext{13}{Gunma Astronomical Observatory, Takayama, Gunma 377-0702, Japan}
\altaffiltext{14}{Earth and Planetary Sciences, School of Science,  Hokkaido University, Kita 10 Nishi8, Kita-ku, Sapporo 060-0810, Japan}
\altaffiltext{15}{Astronomical Institute, Graduate School of Science, Tohoku University, 
6-3 Aramaki Aoba, Aoba-ku, Sendai, Miyagi, 980-8578, Japan}
\altaffiltext{16}{Kiso Observatory, Institute of Astronomy, Graduate School of Science,
The University of Tokyo, Mitake, Kiso-machi, Kiso, Nagano 397-0101, Japan}
\altaffiltext{17}{Faculty of Science, Hokkaido University, Kita 10 Nishi8, Kita-ku, Sapporo 060-0810, Japan}
\altaffiltext{18}{Code 667, NASA Goddard Space Flight Center, Greenbelt, MD 20771, USA}
\altaffiltext{19}{Hiroshima Astrophysical Science Center, Hiroshima University, Higashi-Hiroshima, Hiroshima 739-8526, Japan}
\altaffiltext{20}{Institute for Space-Earth Environmental Research, Nagoya University, 
Furocho, Chikusa-ku, Nagoya, 464-8601, Japan}
\altaffiltext{21}{Okayama Astrophysical Observatory, National Astronomical Observatory of Japan, National Institutes of Natural Sciences, 3037-5 Honjo, Kamogata, Asakuchi, Okayama 719-0232, Japan}
\altaffiltext{22}{Faculty of Education, Saitama University, 255 Shimo-Okubo, Sakura, Saitama, 338-8570, Japan}
\altaffiltext{23}{Department of Physics, Tokyo Institute of Technology, 2-12-1 Ookayama, Meguro-ku, Tokyo 152-8551, Japan}
\altaffiltext{24}{Department of Applied Physics, Okayama University of Science, 1-1, Ridai-cho, Kita-ku, Okayama, Okayama 700-0005, Japan}

%\altaffiltext{2}{Kwasan Observatory, Kyoto University, 
%17-1 Kitakazan-ohmine-cho, Yamashina-ku, Kyoto, 607-8471, Japan} 
%\altaffiltext{5}{Koyama Astronomical Observatory, Kyoto Sangyo University, Motoyama, Kamigamo, Kita-Ku, Kyoto-City 
%603-8555, Japan}
%
%\altaffiltext{9}{Hiroshima Astrophysical Science Center, Hiroshima University, Higashi-Hiroshima, Hiroshima 739-8526, Japan}
%
%\altaffiltext{11}{Subaru Telescope, National Astronomical Observatory of Japan, 650 North
%A'ohoku Place, Hilo, HI, 96720, USA}
%\altaffiltext{12}{Millennium Institute of Astrophysics, Casilla 36-D, Santiago, Chile}
%\altaffiltext{13}{Departamento de Astronom\'ia, Universidad de Chile, Casilla 36-D,
%Santiago, Chile}
%

%
%
%
%
%
%

%\address{}

\begin{abstract}
 We present the early-phase spectra and the light curves
of the broad-lined supernova (SN) 2016coi from $t=7$ to $67$ days 
after the estimated explosion date. This SN was initially reported 
as a broad-lined Type SN Ic (SN Ic-BL). 
However, we found that spectra up to $t=12$ days exhibited the 
He~{\sc i} $\lambda$5876, $\lambda$6678, and $\lambda$7065
absorption lines.
We show that the smoothed and blueshifted spectra of normal SNe Ib
are remarkably similar to the observed spectrum of SN 2016coi.
The line velocities of SN 2016coi were similar to those of
SNe Ic-BL and significantly faster than those of SNe Ib.
%The light curve and color evolution were also similar to 
%those of SNe Ib. 
Analyses of the line velocity and light curve 
suggest that the kinetic energy and the total ejecta mass of SN 2016coi
are similar to those of SNe Ic-BL. 
Together with broad-lined SNe 2009bb and 2012ap for which the detection
of He~{\sc i} were also reported, 
%The ejected $^{56}$Ni mass was smaller than
%those of SNe Ic-BL, while it was usual among SNe Ib.
these SNe could be transitional objects between SNe Ic-BL and SNe Ib,
and be classified as broad-lined Type `Ib' SNe (SNe `Ib'-BL).
Our work demonstrates the diversity of the outermost layer
in broad-lined SNe, which should be related to the variety
of the evolutionary paths.
\end{abstract}

%Uncomment for PACS numbers title message %\pacs{00.00, 20.00, 42.10} % Keywords required only for MST, PB, PMB, PM, JOA, JOB?
%\vspace{2pc}
%\noindent{\it Keywords}: Article preparation, IOP journals % Uncomment for Submitted to journal title message %\submitto{\JPA} % Comment out if separate title page not required \maketitle

\section{Introduction}

Core collapse supernovae (SNe) are classified as Type Ib
when the spectra exhibit the helium but not the hydrogen,
while SNe are classified as Type Ic
when the spectra exhibit neither helium nor hydrogen, 
\citep{Filippenko1997}.
The absorption lines reflect the compositions
of the outer layers of the SN ejecta and progenitor.
It is known that properties of SNe Ib/c have a large variety.
Some SNe Ib/c show broad absorption features,
and such objects are called broad-lined (BL) SN Ic
\citep[SN Ic-BL; ][]{Valenti2008b}.
The high expansion velocities suggest that 
SNe Ic-BL have a higher energy than normal SNe Ib/c
\citep{Foley2003,Valenti2008b}.
SNe Ic-BL associated with a $\gamma$-ray burst (GRB-SNe)
forms a particularly missing subclass \citep{Galama1998,Iwamoto1998}.
Their kinetic energy is likely to be even
larger than those of other SNe Ic-BL \citep{Nomoto2006}.
The origin of these varieties is, however, not well understood yet.
%SNe Ic-BL without GRBs \citep{Foley2003,Valenti2008b}
%are likely to have an intermediate mass and energy
%between normal SNe Ib/c and GRB-SNe.

In theoretical stellar evolution models,
it is not straightforward to produce SN progenitors with no He layer.
Stellar wind and binary interaction play important
roles in removing the envelope.
Interestingly, even in the binary models,
which is probably favored over single star models in removing the envelope,
the progenitors tend to have some amount of helium layer
\citep{Woosley1995,Yoon2010,Yoon2015}.
Thus, to study the evolutionary paths to GRB-SNe and SNe Ic-BL,
it is important to observationally
probe the presence of He in these classes of SNe.

Due to the relatively low event rate, 
the direct detection of the progenitor of SNe Ic-BL
has never been reported to date.
Thus, it is important to study the composition of the progenitor star
from the early-phase spectra of SNe.
In fact, the presence of the He features in optical and near-infrared (NIR)
spectra of SNe Ic-BL has been suggested for SNe 2009bb and 2012ap 
\citep[][]{Pignata2011,Bufano2012,Milisavljevic2015b}.
For SN 2009bb, the identification is, however, based mainly on the 
He {\sc i}~$\lambda$5876 line, which could be contaminated by
the Na {\sc i}~D line \citep{Pignata2011}.
In the spectra of SN 2009bb,
the He~{\sc i}$\lambda$6678 and $\lambda$7065 ones were very weak.
For SN 2012ap, the strong He~{\sc i} $\lambda$10830 line was detected,
but the He~{\sc i}$\lambda$20587 line, which is expected to have
a similar strength to the He~{\sc i} $\lambda$10830,
was weak \citep{Milisavljevic2015b}. 
Also, a statistic analysis of spectra 
of SNe Ic and SNe Ic-BL has been done by \citet{Modjaz2015}.
They reported that these types of SNe
do not show detetacble helium absorption lines.
In sum, the presence of the He layer in SNe Ic and Ic-BL
is still controversial.

In this paper, we present results of our observations of SN 2016coi.
SN 2016coi was discovered at the outskirt of the nearby faint galaxy 
UGC 11868 on May 27.5 by ASAS-SN team \citep{Holoien2016}.
The name of this SN was independently given as ASASSN-16fp.
The distance of the host galaxy was obtained to be
17.2 Mpc using Tully-Fisher relation \footnote{The distance was taken from 
NASA/IPAC EXTRAGALACTIC DATABASE (NED), https://ned.ipac.caltech.edu/}. 
Follow-up spectroscopic observations were performed by several groups
and the spectra were very similar to those of SNe Ic-BL
at their early phase \citep{Elias-Rosa2016}. 
%We just started the ultraviolet-optical-infrared 
%observations in the framework of the ToO program of 
%OISTER \citep[see ][]{Yamanaka2015,Yamanaka2016a}. 

%  In this paper,
We first describe our observations and data reduction in \S 2.
In \S 3, we compare the properties with those of SNe Ib and Ic-BL.
We show that the spectral properties of SN 2016coi
were very similar to those of SNe Ic-BL
except for the detection of the He~{\sc i} absorption lines.
This suggests that SN 2016coi is a transitional SN between
SNe Ib and Ic-BL, and could be classified as an SN `Ib'-BL. 
Finally, we discuss the explosion properties
and progenitor nature of SN 2016coi in \S 4.

%in \S 3.1. %We demonstrate that the early-phase spectra were well similar to those of
%smoothed and blueshifed profile of an SN Ib in \S 3.2. 
%The overall spectral profile was explained by fitting the synthetic spectrum 
%in \S 3.2.
%Finally, we discuss the explosion properties and progenitor nature of 
%SN 2016coi.
%The spectra obtained on May 31, and Jun 1 exhibited the possible 
%strong helium absorption lines \citep{Yamanaka2016b}. 

\begin{figure*}
  \begin{center}
%    \vspace{-1.0cm}
    \begin{tabular}{c}
%      \resizebox{90mm}{!}{\includegraphics{PSNinNGC3447_LC_50.eps}} 
  \resizebox{160mm}{!}{\includegraphics{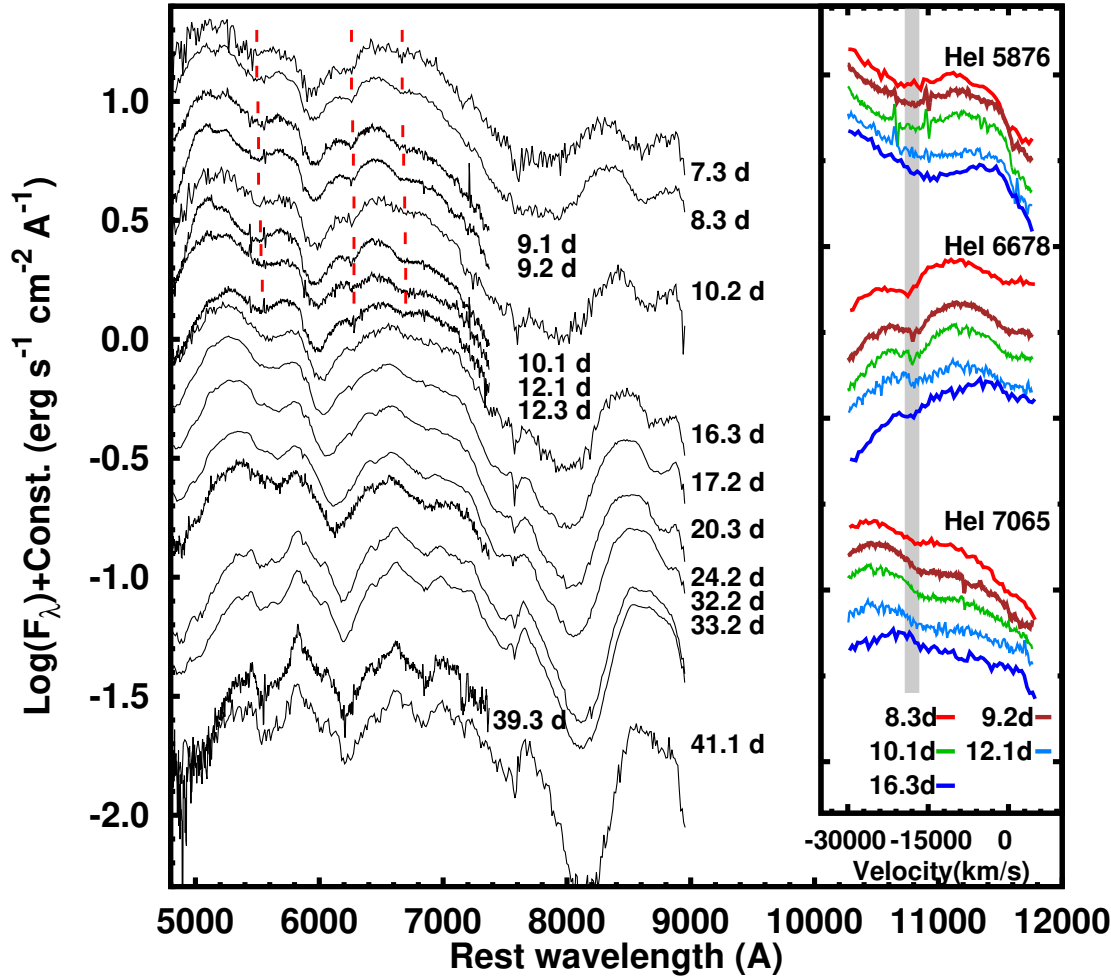}} \\
    \end{tabular}
    \caption{Spectral evolution of SN 2016coi from $t=7$ to $41$ d. 
 Wavelengths of spectra were corrected for using the recession velocity of 
 the host galaxy UGC 11868 (z=0.0036; \cite{Giovanelli1993}).
 Spectra were corrected for the atmospheric lines using the spectrophotometric 
 standard star spectra. 
 (Right inset) The close-up view of the 
 He~{\sc i}$\lambda$5876, $\lambda$6678, and $\lambda$7065 regions
 shwon in the Doppler velocity.
 The vertical gray lines show the velocity of 18,000 km~s$^{-1}$.}
    \label{lc}
  \end{center}
\end{figure*}

\begin{figure}
  \begin{center}
%    \vspace{-1.0cm}
    \begin{tabular}{c}
%      \resizebox{90mm}{!}{\includegraphics{PSNinNGC3447_LC_50.eps}} 
 \resizebox{90mm}{!}{\includegraphics{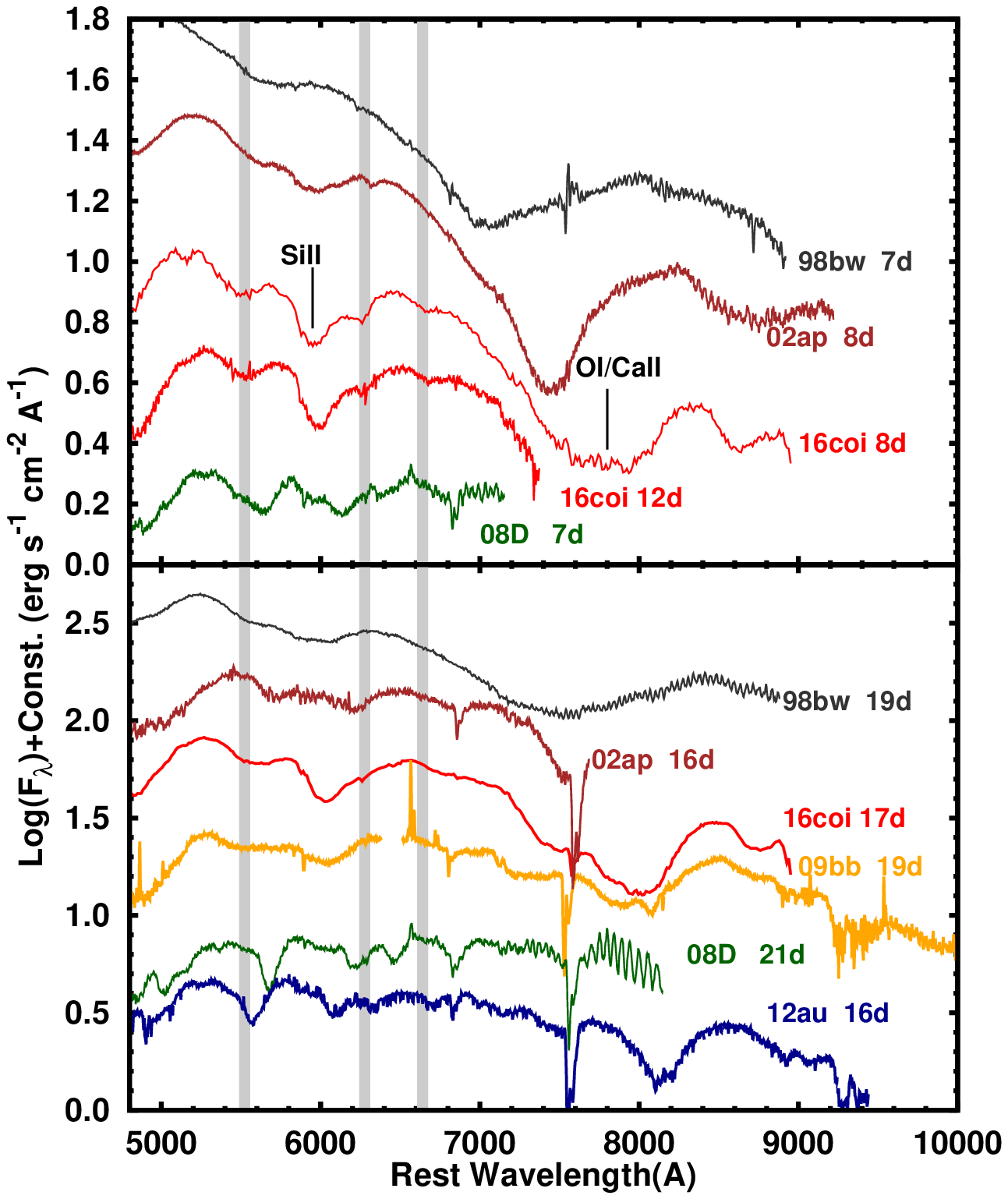}} \\
    \end{tabular}
    
    \caption{
      (Top panel) Spectra of SN 2016coi at $t=8$ and $12$ d
      compared with those of SNe Ic-BL 1998bw
\citep{Patat2001}, 2002ap \citep{Foley2003}, and SN Ib 2008D 
\citep{Modjaz2009}. These data were taken from
 the SUSPECT \footnote{https://www.nhn.ou.edu/~suspect/} and WISEREP 
 \footnote{http://wiserep.weizmann.ac.il/; \citep{Yaron2012}} databases.
 Vertical grey lines denote the wavelength the He~{\sc i} $\lambda$5876, 
 $\lambda$6678,  and $\lambda$7065 which were blueshifted with
 18,000 km~s$^{-1}$.
 (Bottom panel) The spectrum at $t=17$ d compared with those of
 other SNe as in the top panel as well as
 SNe 2009bb \citep{Pignata2011} and 2012au \citep{Takaki2013}.
}
\label{lc}
\end{center}
\end{figure}

\section{Observations and Data Reduction}

 Observations of SN 2016coi were performed using various telescopes and 
instruments in the framework of the Target-of-Opportunity (ToO) program in 
the Optical and Infrared Synergetic Telescopes for Education and Research
(OISTER).
 Optical spectroscopic observations were performed using the 1.5-m Kanata 
telescope attached with the Hiroshima One-shot Wide-field Polarimetry 
\citep[HOWPol; ][]{Kawabata2008} on 14 nights from May 31 through Jul 4.
Spectroscopic observations were also performed using the 1.5-m telescope
attached with the Gunma
LOW-resolution Spectrograph and imager (GLOWS) on three nights 
from Jun 2 through Jun 17, and the 2.0-m Nayuta telescope attached with 
the Medium And Low-dispersion Long-slit Spectrograph (MALLS) on 
four nights from Jun 2 through Jul 2. 
  
 Spectroscopic data were reduced according to the standard manner.
Wavelength calibrations were performed for the 
HOWPol data using atmospheric lines observed in the same frame to 
the object, and for the GLOWS and the MALLS using the FeNeAr lamp. 
Flux calibrations were performed using the bright high-temperature
standard stars.
The atmospheric lines of the SN spectra were corrected 
for using the standard-star spectra.
%(see top of panel in Figure 1).

 Optical imaging observations were performed using the Kanata telescope 
attached with the HOWPol on 17 nights from May 31 through Jul 26, 
the 1.6-m Pirka telescope attached with the Multispectral Imager 
\citep[MSI; ][]{Watanabe2012} on 21 nights 
from Jun 2 through Jul 25, the 1.05-m telescope attached with the Kiso Wide Field
Camera \citep[KWFC; ][]{Sako2012} on six nights from Jun 2 through Jun 17, 
and the 1.05-m telescope in the Multicolor Imaging Telescopes for Survey and Monstrous 
Explosions \citep[MITSuME; ][]{Kotani2005} at the Ishigakijima 
Astronomical Observatory on 28 nights from Jun 3 through Jul 30. 
Imaging observations of photometric standard star fields were also
performed using the HOWPol and the MSI on photometric nights.
Near-infrared photometry were performed
using the 1.4-m InfraRed Survey Facility 
(IRSF) telescope attached with the Simultaneous three-color InfraRed Imager
 for Unbiased Survey \citep[SIRIUS; ][]{Nagayama2003} 
on Jun 3, the Kanata telescope 
attached with the Hiroshima Optical and Near-InfraRed camera 
\citep[HONIR; ][]{Sakimoto2012,Akitaya2014,Ui2014} on Jun 1, 
the Nayuta telescope attached with Nishi-harima Infrared Camera (NIC) 
on Jun 14, and the 1.0-m telescope at Kagoshima University on four nights 
from Jun 14 through Jul 21. 
 
 Reductions of the imaging data were performed 
in the similar methods to the previous studies 
\citep[see ][]{Yamanaka2015,Yamanaka2016a}.
The point spread function (PSF)
photometry were performed using the $IRAF$ software $DAOPHOT$.
Standard magnitudes
of the local standard stars were obtained using the 
standard star magnitudes \citep{Landolt1992}. 
Systematic differences among different instruments were confirmed. 
We corrected for the color terms of HOWPol \citep{Kawabata2008}, 
MSI \citep{Watanabe2012}, KWFC \citep{Sako2012}, and MITSuME \citep{Kotani2005}
when SN magnitudes were calculated.
The color terms of these instruments were already obtained using 
the secondary standard stars in M67 \citep[see also][]{Yamanaka2015,Yamanaka2016a}. 
The systematic errors were calculated using the average of the 
residual differences. 

 The extinctions were corrected only for 
 Galactic dust \citep{Schlafly2011}.
 This is supported by the fact that 
 the sodium absorption lines only from our Galaxy were
 detected in the spectra.
 The total extinction of $A_{V}=0.2$ \citep{Schlafly2011}
 is adopted throughout this paper. 
 The extinction coefficient is assumed to be $R_{V}=3.1$ as a typical value.

 \section{Results}

    \subsection{Spectral features}

\begin{figure}
  \begin{center}
     \begin{tabular}{c}
%      \resizebox{90mm}{!}{\includegraphics{PSNinNGC3447_LC_50.eps}} 
 \resizebox{85mm}{!}{\includegraphics{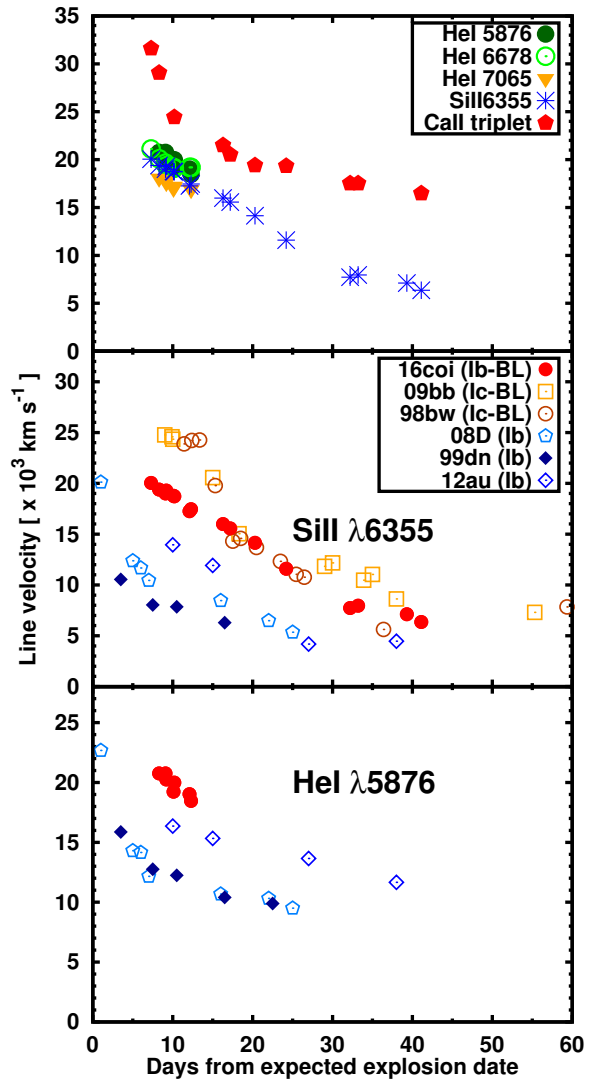}} \\
    \end{tabular}
    \caption{(Top panel) Velocity evolutions of Si~{\sc ii}$\lambda$6355, 
 Ca~{\sc ii}~IR~triplet, He~{\sc i}$\lambda$7065, $\lambda$6678, and 
 $\lambda$5876 lines.
 These line velocities were estimated from the absorption minimum
 measured using the $splot$ task equipped with $IRAF$.
 (Middle panel) The line velocities of the Si~{\sc ii} $\lambda$6355 line
 compared with those of SNe Ic-BL 1998bw, 2009bb, SNe Ib 1999dn, 2008D, and 2012au.
 (Bottom panel) Similar to the middle panel, but
 for the He~{\sc i}~$\lambda$5876 line.}
    \label{lc}
  \end{center}
\end{figure}

 %The line velocity of 
 %SN Ib 1999dn was added for this comparison \citep{Benetti2011}.

\begin{figure*}
  \begin{center}
    \begin{tabular}{c}
%      \resizebox{90mm}{!}{\includegraphics{PSNinNGC3447_LC_50.eps}} 
\resizebox{140mm}{!}{\includegraphics{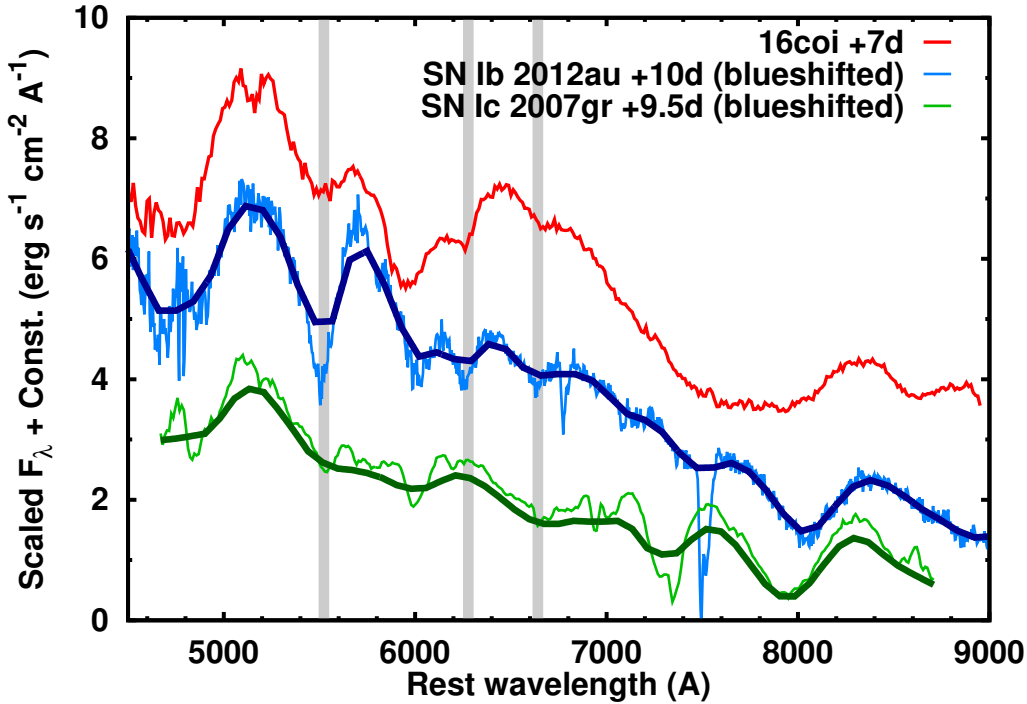}} \\
    \end{tabular}
    \caption{The spectrum of SN 2016coi at $t=7$ d compared with the
      smoothed and blueshifted spectra of SN Ic 2007gr \citep{Yamanaka2016c}
      and SN Ib 2012au. The flux intensity of these two SNe were artificially 
      scaled for the comparison.
      The degree of blueshift is determined to match the 
      He {\sc i} absorption lines of SN 2016coi.
      Gaussian kernel smoothing were also performed for these two SNe
      to match the width of He~{\sc i} absorption lines of SN 2016coi.
      The vertical lines denote the He~{\sc i} absorption lines
      at their velocity of 18,000 km~s$^{-1}$. }
    \label{lc}
  \end{center}
\end{figure*}

\begin{figure*}
  \begin{center}
    \begin{tabular}{c}
%      \resizebox{90mm}{!}{\includegraphics{PSNinNGC3447_LC_50.eps}} 
 \resizebox{140mm}{!}{\includegraphics{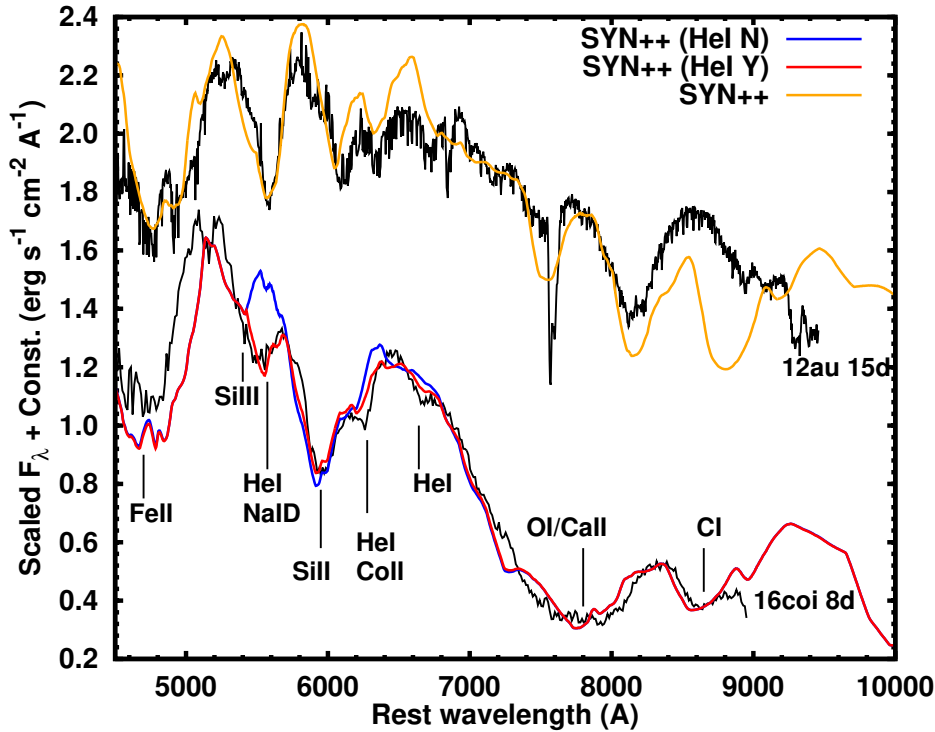}} \\
    \end{tabular}
    \caption{The observed spectrum of SN 2016coi at $t=8$ d compared with 
 the synthetic spectrum constructed using the SYN++ \citep{Thomas2011}. 
 The He~{\sc i}, C~{\sc i}, O~{\sc i}, Na~{\sc i},  
 Si~{\sc ii}, Si~{\sc iii}, Ca~{\sc ii}, Fe~{\sc ii} and Co~{\sc ii}
 features were used to calculate the spectrum.
 The synthetic spectrum without the only He~{\sc i} 
 feature was also presented.}
    \label{lc}
  \end{center}
\end{figure*}

  Figure 1 shows the spectral evolution of SN 2016coi
  from $t=7$ to $41$ d
  \footnote{$t=0$ is the estimated explosion date and defined as MJD 57532.5. 
    See \S 3.3 for details.} . The corrected redshift is z=0.0036 for SN 2016coi.
  The spectra exhibited the strong absorption lines
 around 6000 \AA. The feature was identified as Si{\sc ii} $\lambda$6355
 and its velocity reached 18,000 km~s$^{-1}$. The broad absorption line
 around at 8000 \AA\ was blended by the O{\sc i}~$\lambda$7774 and 
 Ca{\sc ii} IR triplet due to their extremely high velocities.
 Features marked by the red lines were attributed to 
 He~{\sc i}. Identification of He~{\sc i} will be discussed in \S 3.2.

  The spectrum at $t=8$ d was compared with those of 
 SNe Ic-BL 1998bw \citep{Patat2001}, 2002ap \citep{Kawabata2002,Foley2003}, 
 and SN Ib 2008D \citep{Modjaz2009} (see the top panel of Figure 2). 
 The intensity of
 Si~{\sc ii} $\lambda$6355 absorption line of SN 2016coi was significantly 
 stronger than those of other SNe. Although the absorption depth generally 
 becomes shallower when the feature is broadened as seen in SN 1998bw,
 the absorption line of SN 2016coi was deep and broad.

  The broad feature around 7400--8200 \AA\ 
 could not be separated into different lines. 
 This characteristics is common in SNe Ic-BL.
 For SN 1998bw, the Ca~{\sc ii} IR triplet was
 blueshifted to $\sim$7000 \AA\
 due to its extremely high-velocity ejecta and the feature was even 
 blended with O~{\sc i} $\lambda$7774
 \citep{Tanaka2007}. For SN 2002ap, the feature was dominated 
 by the O~{\sc i} only \citep{Kawabata2002,Foley2003}.
% because the calcium was not so dominant in its ejecta 
   
 For the comparison of the spectrum at $t=17$ d,
 the spectra of SN Ic-BL 2009bb \citep{Pignata2011}
 and SN Ib 2012au \citep{Takaki2013} were added
 (see the bottom panel of Figure 2). 
 The line velocity of Si~{\sc ii}~$\lambda$6355 was similar
 to those of SNe Ic-BL 1998bw and 2009bb 
 (see also the middle panel of Figure 3), but significantly faster than 
 those of SNe Ib 2008D and 2012au.
 The broad feature composed of the O~{\sc i} and Ca~{\sc ii} lines
 was still seen at this epoch.
 This feature was also similar to those of SNe 1998bw and 2009bb.
 The same feature in SN 2012au was 
 completely separated into the O~{\sc i} and Ca~{\sc ii}.
 In summary, SN 2016coi is similar to an SN Ic-BL in many respects.
 However, as discussed below, 
 the strong He~{\sc i} features were detected and the intensities
 were anomalously strong for SNe Ic-BL. 
 With the spectra exhibiting He~{\sc i}, this SN should be
 classified as an SN `Ib'-BL according to the definition.
% These facts mean that

 \subsection{Detection of He~{\sc i}}
  Spectra up to $t=12$ d exhibited the absorption lines at 5500, 
 6300, and 6850 \AA\ (see the inset of Figure 1). 
 In this section, we show that these features were attributed 
 to the absorption lines of 
 He~{\sc i} $\lambda$5876, $\lambda$6678, and $\lambda$7065. 
 The velocities reached 18,000 km~s$^{-1}$,
 and the corresponding wavelengths were denoted by the vertical 
 gray lines in Figures 1 and 2. 
% These similar velocities indicate 
% that He~{\sc i} absorption lines were correctly identified.

 In order to confirm the line identification of He~{\sc i}, 
 we first produced artificially smoothed spectra of SN Ic 2007gr
 \citep{Yamanaka2016c} and SN Ib 2012au
 (see Figure 4). 
 We smoothed the spectra by using Gaussian kernel
 so that the Full-width-of-Half-Maximum (FWHM)
 of absorption lines of SN 2016coi (11,000 km~s$^{-1}$) matches to 
 those of SNe 2007gr (5,000 km~s$^{-1}$) and 2012au (7,500 km~s$^{-1}$).
 Then, the spectra were further blueshifted to match the positions 
 of the absorption lines.
 As shown in Figure 4,
 the spectrum of SN 2016coi was very similar to the smoothed and
 blueshifted spectrum of SN 2012au.
 We confirmed that
 in addition to the He~{\sc i}~$\lambda$5876 line, 
 both the He~{\sc i}~$\lambda$6678 and $\lambda$7065 
 features were similar in the spectrum of SN 2016coi
 and the smoothed and blueshifted spectrum of SN 2012au.
 On the other hand, as expected from Type Ic identification, 
 the smoothed and blueshifted spectrum of SN 2007gr did not exhibit
 the He~{\sc i} absorption lines.
 These tests demonstrate that SN 2016coi show the He lines.

% the spectrum of SN 2016coi was very similar to the smoothed and
% blueshifted spectrum of SN 2012au.Remarkably,

%  The smoothed spectrum of SN 2016coi was also constructed to compare
% with the spectrum of SN 1998bw (see the bottom panel of Figure 4).  
% The smoothed spectrum of SN 2016coi still exhibited the 
% broadened features of the He~{\sc i}~$\lambda$5876 absorption lines, 
% but not He~{\sc i}~$\lambda\lambda$6678,7065.
% Any He~{\sc i} features were not found in SN 1998bw. 
% The strong peak around 6200 \AA\ 
% was also found for SN 2016coi, but not for SN 1998bw. 
% These facts support that the He~{\sc i}
% features would not be found in the SN 1998bw-like SNe 
% \citep[see also][]{Modjaz2015}. 
% However, the synthetic spectrum, 
% SYN++ \citep{Thomas2011} composed of only He~{\sc i} absorption lines 
% cannot explain the intensity ratio of the features.
 
  The relative absorption depths of the He lines were
 slightly different in SNe 2016coi and 2012au.
 We thus further study the possible He~{\sc i} features of SN 2016coi
 using the synthetic spectral code, SYN++ to assess possible contamination 
 of the other lines in the He features. 
\citep{Thomas2011}. 
 The O~{\sc i}, Na~{\sc i}, Si~{\sc ii}, Si~{\sc iii}, Ca~{\sc ii}, Fe~{\sc ii}, 
 and Co~{\sc ii} were used to reproduce the depth ratio.
 %Interestingly,
 %the strength ratio of the features of SN 2016coi
 %cannot be explained by only He~{\sc i}. 
 %The synthetic spectrum exhibits that the He~{\sc i}$\lambda$5876
 %was too strong and the He~{\sc i}6678 was too weak
 %to explain the observed spectrum. 
 %The possible other features could be needed to explain their line strength.
 %The Na~{\sc i}, Si~{\sc iii}, and Co~{\sc ii} lines were added to 
 %the synthetic spectrum (see Figure 5).
 Figure 5 shows the Na~{\sc i}D and Si~{\sc iii} features
 would contaminate to the broad line at 5600 \AA.
 The Co~{\sc ii}~$\lambda\lambda$5914,6019 help to explain the strength of
 the feature at 5600 \AA. The Co~{\sc ii}~$\lambda\lambda$6540,6570 
 also support to explain the strong absorption line at 6200\AA\ (see Figure 5). 
 The absorption lines 
 at 5600, 6200 and 6600\AA\ were well explained using these 
 components. On the other hand, it is hard to explain the depth ratios 
 without He~{\sc i} lines (see Figure 5).   
% {\bf (MT: this is not enough to show you really need He.
%   Show the models with/without He)}

% the absorption line at 4800\AA\, but the strong emission was at seen 5000\AA.
% The Fe~{\sc ii} was needed to explain
% The absorption feature of He~{\sc i}$\lambda$6678 was significantly stronger than
% that indicated by the synthetic spectrum, while the He~{\sc i}$\lambda$5876 
% was weaker. This trend was inconsistent with that of SN 2012au 
% \citep{Milisavljevic2013}, which showed that the He~{\sc i}$\lambda$7065 is 
% the strongest feature among three lines. 
 
% The Co~{\sc ii} feature exhibits the emission component 
% at 5500\AA\, while it exhibits the absorption lines at 6300\AA.
% This suggets that the Co~{\sc ii} 
% features could support to explain the anomalous intensity ratio of these 
% absorption lines. 
%  The intensity ratio was 
% also naturally explained by including the Co~{\sc ii} lines. Inversely, the 
% absorption line ratio cannot be explained by only Co~{\sc ii}, suggesting that 
% the identification of He~{\sc i} lines was real for SN 2016coi.
% The stronger absorption line at 6400\AA for SN 2016coi than SN 2012au 
% means that the ejecta may contain larger cobalt mass, implying that the 
% explosion energy may be related.

 Note that inclusion of Co~{\sc ii} at 3800 \AA, 4400 \AA, and other 
 elements is also consistent with a blue portion of the spectrum.
 Since Fe-group elements strongly affect the spectra
 in the ultraviolet wavelengths,
 we calculated $U-B$ color in the synthetic spectrum.
 The calculated color ($U-B$ =0.6 mag)
 is consistent with observed $U-B$ within the error.  
 The Mg~{\sc ii} and Ti~{\sc ii} features possibly contributes to
 the spectra below 4500 \AA,
 but again the U-band flux was consistent with the observed color
 even including these features.
 %contribution from the constructed spectrum sensitively 
 %changes when the strengths of absorption lines of iron-group elements 
 %are changed.

 To study the validity of our spectral fit, we 
 also fit a spectrum of SN 2012au using the same ion species
 (He~{\sc i}, C~{\sc i}, O~{\sc i}, Na~{\sc i}, Si~{\sc ii}, Si~{\sc iii}, 
 Ca~{\sc ii}, Fe~{\sc ii}, and Co~{\sc ii}).
 The spectrum could be explained by the synthetic one except for 
 slight differences at 6200 and 6600 \AA. The absorption features 
 of He~{\sc i} were well matched. This suggests that the Co~{\sc ii} 
 features could support to improve the spectral fit. 

% There may be any similarities with objects like SNe 2009bb and
% 2012ap, and the  should be also encouraged in the future. 

%  \footnote{ SNe 2009bb and 2012ap were pointed out having common properties 
% related to the central engine activity 
% from the multi-wavelength observations 
% \citep{Soderberg2010,Margutti2014,Chakraborti2015,Nakauchi2015}.
% These are beyond the scope for this Letter.}
% Synthetic spectral code were used for 
% the identification of He~{\sc i} of these SNe, because their possible 
% helium features were very weak.

 In summary, we identified the He features in the pre-maximum spectra 
 of SN 2016coi.
 The detection of the He lines has been suggested in other SNe Ic-BL.
 For SN Ic-BL 2009bb, the identification of He~{\sc i} 
 has been discussed \citep{Pignata2011}.
 In SN 2009bb, the He~{\sc i}$\lambda$6678 and He~{\sc i}$\lambda$7065 features
 were very weak and the identification was marginal
 (see Figure 2).
 The near-infrared spectrum for SN 2012ap exhibited a
 strong absorption line around 10500 \AA\ \citep{Milisavljevic2015b}.
 The NIR and optical features were simultaneously explained using the 
 synthetic spectra, but they pointed out that the 
 He~{\sc i}$\lambda$20587 feature was too weak to explain the strength 
 of He~{\sc i}$\lambda$10830 \citep[see also ][]{Bufano2012}. 
 Compared with these two previous cases,
 early-phase spectra of SN 2016coi gives more robust
 identification of the He~{\sc i} lines, because
 the lines were stronger than those of SN 2009bb and 2012ap. 
 
 The bottom panel of Figure 3 shows the 
 comparison of the He~{\sc i}$\lambda$5876 line velocities with 
 those of SNe Ib 1999dn \citep{Benetti2011}, 2008D, and 2012au.
 The He velocity of SN 2016coi 
 was almost identical to that of Si~{\sc ii}$\lambda$6355,
 and was the fastest among other SNe Ib until $t=12$ d.
 Again, the Si~{\sc ii}$\lambda$6355 line velocity of SN 2016coi
 was very similar to those of SNe 1998bw and 2009bb 
 after $t=15$ d. These facts support that
 SN 2016coi could be classified as an SN `Ib'-BL.
 %SNe 2009bb and 2012ap may be members of SNe `Ib'-BL.

\subsection{Light curves}

\begin{figure}
  \begin{center}
    \begin{tabular}{c}
%      \resizebox{90mm}{!}{\includegraphics{PSNinNGC3447_LC_50.eps}} 
 \resizebox{85mm}{!}{\includegraphics{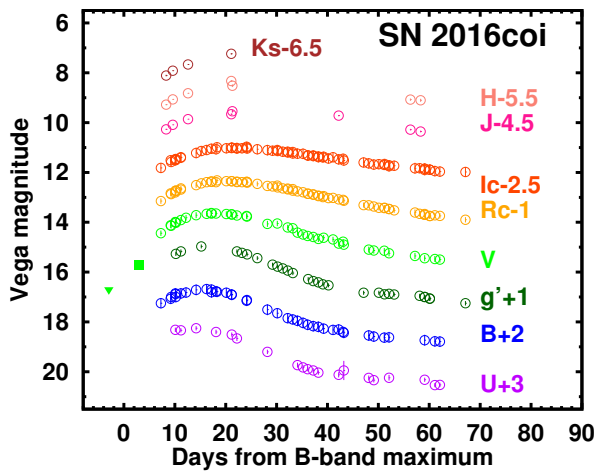}} \\
    \end{tabular}
    \caption{$UBg'VRIJHKs$-band light curves of SN 2016coi.
 The explosion date was estimated to be MJD 57532.5 ($t=0$) 
 by extrapolating the rising part.
 The triangles denote the upper limit reported in \citet{Holoien2016}. }
     \label{lc}
  \end{center}
\end{figure}

\begin{figure}
  \begin{center}
    \begin{tabular}{c}
%      \resizebox{90mm}{!}{\includegraphics{PSNinNGC3447_LC_50.eps}} 
\resizebox{85mm}{!}{\includegraphics{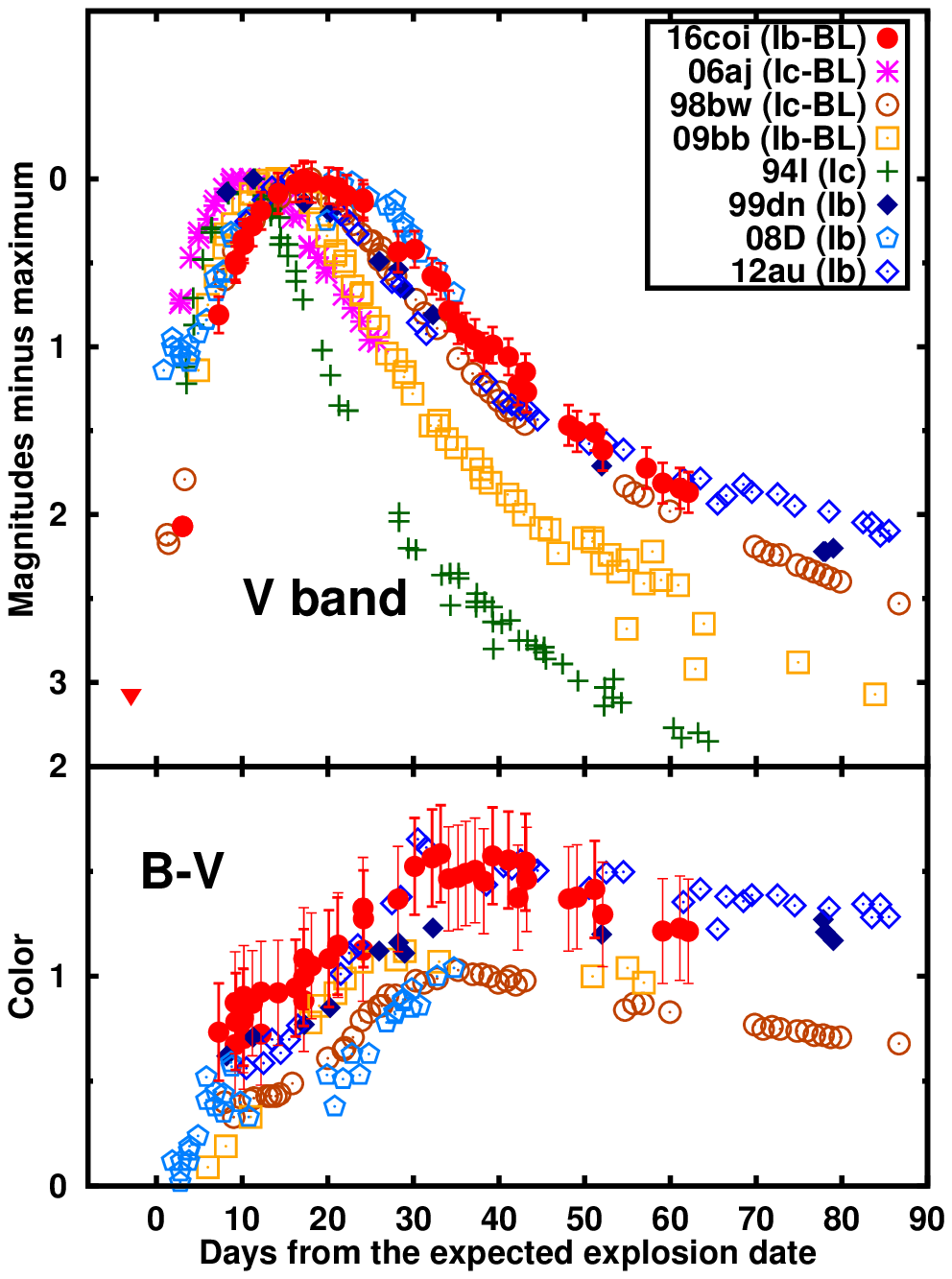}} \\
    \end{tabular}
    \caption{ (Top panel) The $V$-band light curve 
 compared with those of SNe Ic-BL 1998bw 
 \citep{Clocchiatti2011}, 2006aj 
 \citep{Ferrero2006}, 2009bb \citep{Pignata2011}, a normal SN Ic 1994I
 \citep{Richmond1996}, SNe Ib 2008D \citep{Modjaz2009}, and 
 2012au \citep{Takaki2013}.
The light curves are shown in the magnitudes relative to the maximum.
 (Bottom panel) The $B-V$ color evolution compared with those of 
other SNe as in the top panel.
 Color excesses due to the Galactic and host galactic 
 extinction were corrected for using the reported values in each reference. 
 }
     \label{lc}
  \end{center}
\end{figure}

\begin{figure}
  \begin{center}
    \begin{tabular}{c}
%      \resizebox{90mm}{!}{\includegraphics{PSNinNGC3447_LC_50.eps}} 
\resizebox{85mm}{!}{\includegraphics{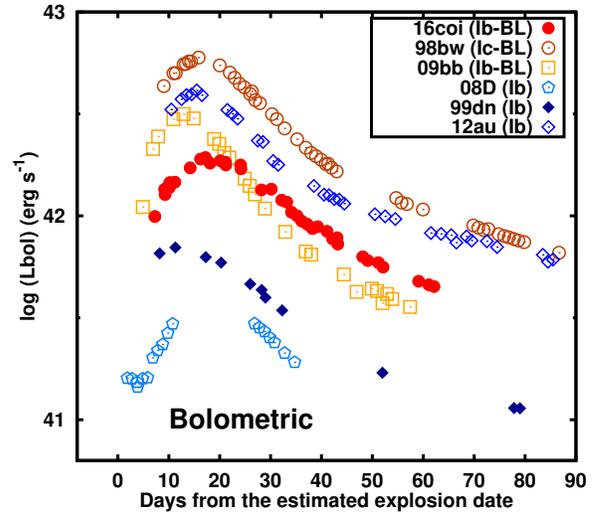}} \\
    \end{tabular}
    \caption{ Quasi-bolometric light curve compared with those of other SNe.
      The bolometric light curves were constructed
      using optical ($BVRI$-band) light curves
      assuming that the optical fluxes are 60\% to the total luminosity. 
 }
     \label{lc}
  \end{center}
\end{figure}

  Figure 6 shows the multi-band light curves of SN 2016coi.
  The $V$-band light curve showed the fast rise up
  in the first two data points. 
By extrapolating the rising part, the explosion date was estimated to be MJD 57532.5 (May 24.5).
 This is consistent with the upper-limit magnitude reported 
 by \citet{Holoien2016}, and this date is denoted 
 to be $t=0$ throughout this paper.
 The $V$-band light curves reached the maximum magnitude at $t=17$ d. 

%  The light curve was compared with those of other well-observed SNe Ib/c,
% e.g., SNe Ic-BL 1998bw \citep{Clocchiatti2011}, 2006aj 
% \citep{Ferrero2006}, 2009bb \citep{Pignata2011}, a normal SN Ic 1994I
% \citep{Richmond1996} and SNe Ib 1999dn \citep{Benetti2011}, 
 %2008D \citep{Modjaz2009}, 2012au \citep{Takaki2013} (see the second panel
%of Figure 4).
 The rising part of the light curves exhibited the slow evolution
 which was similar to those of SNe 1998bw, 2008D, and 2012au
 (see the upper panel of Figure 7). 
 The estimated rise time of 17 days is almost similar to
 16.5 days of SN 2012au and 18 days of SN 1998bw, and slightly shorter 
 than that of SN 2009bb.
 SN 2008D has a longer rise time than that of SN 2016coi.
 The decline rate of SN 2016coi was also similar to 
 those of SNe 1998bw, 1999dn, 2008D, and 2012au,
 but significantly slower than those of SNe 1994I and 2009bb.

 The bottom panel of Figure 7 shows $B-V$ color evolution.
 The color excesses were corrected for using the values
 reported in each reference. 
 The $B-V$ evolution of SN 2016coi was 
 very similar to those of SNe 1999dn and 2012au,
 but different from that of SNe 1998bw, 2008D, and 2009bb.
 The color was always redder than those of SNe Ic-BL,
 and rather similar to those of SNe Ib.

%  \subsection{Bolometric luminosity and $^{56}$Ni mass}

 To derive the absolute bolometric luminosity of SN 2016coi,
 the integration of the muli-band light curve was performed.
 The contribution of the optical emission 
 to the total bolometric luminosity around the maximum date
 was assumed to be $60\%$.
 The distance modulus of $\mu=31.18$ mag was adopted,
 which was taken from $NED$.
 For the comparison,
 the quasi-bolometric light curves of SNe 1998bw, 
 1999dn, 2008D, 2009bb and 2012au were also constructed
 in the same manner.
 Figure 8 shows the quasi-bolometric light curves of these SNe.
 
 The peak luminosity of SN 2016coi was estimated to
 be $3.2\times10^{42}$ erg~s$^{-1}$.
 It was substantially fainter than those of SNe 1998bw, 2009bb and 
 2012au. It was more luminous than those of SNe 1999dn and 2008D.
 The $^{56}$Ni mass of SN 2016coi was estimated to be $\sim0.15 
 M_{\odot}$ assuming that the rise time was 17 days and $\alpha=1$
 \citep{Arnett1982,Stritzinger2005}.
 This $^{56}$Ni mass was relatively small among SNe Ic-BL,
 while it is more consistent with normal SNe Ib.

 %Discussion from here??
 
% The detection of radio emissions were reported by various 
% groups \citep{}. 

\begin{figure}
  \begin{center}
    \begin{tabular}{c}
%      \resizebox{90mm}{!}{\includegraphics{PSNinNGC3447_LC_50.eps}} 
 \resizebox{85mm}{!}{\includegraphics{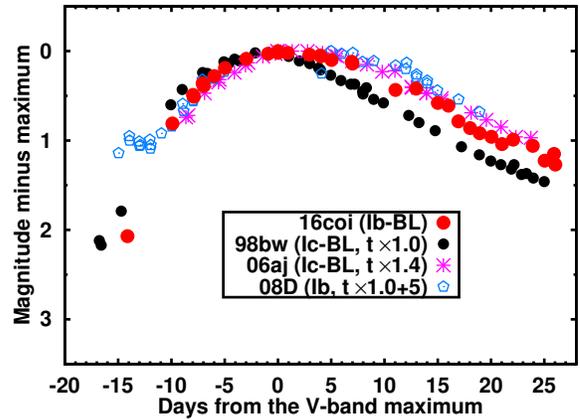}} \\
    \end{tabular}
    \caption{The $V$-band light curves compared with the 
      stretched and shifted light curves of SNe 1998bw, 2006aj, and 2008D.
      }
    \label{lc}
  \end{center}
\end{figure}
  
\section{discussion and conclusion}

 In this paper, we showed results of our photometric and spectroscopic
observations of SN 2016coi.
The He~{\sc i} lines were clearly identified in the spectra.
Except for the presence of He~{\sc i},
the spectral features of SN 2016coi were similar to those of SNe Ic-BL.
The line velocity as well as
broad features of Si~{\sc ii} and Ca~{\sc ii} absorption lines 
were also similar to SNe Ic-BL.
\citet{Pignata2011} pointed out
that the He~{\sc i} features were detected in spectra of SN Ic-BL 2009bb.
The detection of the possible He~{\sc i} features in NIR spectra 
were also reported for SN 2012ap \citep{Milisavljevic2015b}. 
The detection of the helium in these SNe means
that these are classified as SNe Ib according to the definition.
We suggest the classification of these SN to be `Ib'-BL.

We estimate the kinetic energy and the total ejecta mass of SN 2016coi
using the scaling method \citep[see ][]{Sahu2008}. 
% For the analysis, the comparison of the light curve timescales were 
% performed using the pre- and post- maximum ones.  
Since the light curve shape of SN 2016coi was similar
to the stretched light curve of SNe 2006aj and 2008D
(see Figure 9), we use these objects as references.
%{\bf (MT: stretched 06aj and 08D do not look similar to 16coi...)}
% The stretches of the light curves of comparative SNe were performed to 
% match with that of SN 2016coi (see Figure 5). 
The light curve timescale of SN 2016coi 
was 1.4 times longer than that of SN 2006aj
while it is comparable to that of SN 2008D.
The Si~{\sc ii} line velocity of $\sim15,000$ km~s$^{-1}$
was similar to of SN 2006aj and
1.6 times larger than the 9000 km~s$^{-1}$ of SN 2008D.
Thus, by using SN 2006aj as a reference \citep{Mazzali2006},
the total ejected mass and the kinetic energy were estimated to be
$M_{ej}\sim 10M_{\odot}$ and $E_{k}\sim 3.0\times 10^{52}$ erg.
Similarly, by using SN 2008D as a reference \citep{Tanaka2009a}.
the kinetic energy and the total ejected
mass were estimated to be
$M_{ej} \sim 10M_{\odot}$ and $E_{k}\sim5.0\times 10^{52}$ erg.
The estimated ejecta masses and the kinetic energies
were similar to those of GRB-SNe.
% Explosion properties were rather similar to those of SNe Ic-BL.

%{\bf (MT: Do you need this paragraph??)}
%It should be noted that although
%the comparison of the explosion parameters with those
%of SNe 2006aj and 2008D is rather robust,
%it is not robust with SN 1998bw.
%{\bf (MT: I do not understand the following statements)}
%The velocity of SN 1998bw around at $t=10$ d was larger
%than that of SN 2016coi. The light curve and the velocity 
%evolution of SN 2016coi was different from those of SN 1998bw. 
%This implies that the explosion structure of SN 2016coi may be 
%different from that of SN 1998bw.

The presence of He in other broad-lined SNe is of great interest.
To study this possibility,
we further smoothed and blueshifted spectrum of SN 2016coi
to match the absorption velocity and width to those of SN 1998bw
(see Figure 10). 
We confirmed that smoothed spectrum of SN 2016coi still
shows a hint of the He {\sc i} features
while it is not seen in the spectrum of SN 1998bw.
Therefore, SN 1998bw does not have a similar amount of He.
These considerations suggest that SNe 2016coi (also SNe 2009bb and 2012ap) 
 may belong to
a different class of SNe from the prototype of broad-lined SNe.
This is in line with the finding by \citet{Modjaz2015},
who concluded that SNe Ic-BL are He free.
There may be diversity among broad-lined SNe
in terms of helium contents in the outer layer.
A caveat is that it would also be possible that other broad-lined SNe still contain
the He layer, but the He lines are not strong in absence of the
non-thermal excitation \citep[e.g.,][]{Hachinger2012}.

%could not have the helium layer, and completely dissipate the helium.
%However, we demonstrate that the strong He~{\sc i} absorption lines
%were clearly detected for SN 2016coi in this paper.

 This work demonstrates on important role of very early phase 
 observations for the identification of He in broad-lined SNe.
%We here discuss the detectability of the helium features in SNe Ic-BL.
The He~{\sc i} absorption lines seen in SN 2016coi became very weak at $t=15$ d,
while the He~{\sc i} lines generally become stronger
at the later epoch for SNe Ib \citep{Matheson2001}. 
% \footnote{If the progenitor underwent more 
% stripped-envelope than SNe Ib with a small amount of hydrogen, this would be 
% naturally interpreted.} 
We speculate that the He~{\sc i} lines may have eluded the detection
due to the lack of pre-maximum early phase spectra.
%It must be important to follow the evolution of the He {\sc i} lines
%for this SN until late epochs.

%  We presented that SN 2016coi could be classified as an SN `Ib'-BL 
% which has the rich helium in the outer-part ejecta and the large
% kinetic energy. We discuss the explosion property of SN 2016coi and 
% present the matters related to the progenitor which has the helium envelope.
  
%  \subsection{Why was helium found?}
%  Similar propoerties to SNe 1998bw and 2009bb were presented. SN 1998bw is a
% prototype of the highly-energetic SN associated with the GRB 
% \citep[GRB-SN][]{Galama1998,Iwamoto1998}, and
% SN 2009bb is also an energetic event related to the relativistic jet from the 
% central engine \citep{berg2010,Pignata2011}. However, He~{\sc i}
% features were not detected in their early-phase spectra, while For SN 2009bb, 
% the He~{\sc i} was possibly identified at the ??-?? days after the 
%maximum date \citep{Pignata2011}. 

%  The detection of the strong absorption lines of He~{\sc i} in spectra of 
% SNe Ic-BL have never been reported to date except for the several marginal case. 

\begin{figure}
  \begin{center}
    \begin{tabular}{c}
%      \resizebox{90mm}{!}{\includegraphics{PSNinNGC3447_LC_50.eps}} 
\resizebox{85mm}{!}{\includegraphics{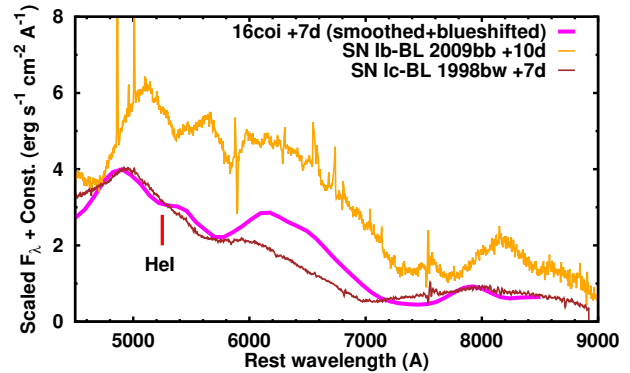}} \\
    \end{tabular}
    \caption{Blueshifted and smoothed spectrum of SN 2016coi compared with 
 spectra of SNe Ic-BL 1998bw and 2009bb. The He~{\sc i} $\lambda$5876 
 feature was seen in SN 2016coi, but not in SN 1998bw.}
    \label{lc}
  \end{center}
\end{figure}

Our work demonstrated a diversity of the outermost layer in broad-lined SNe.
The presence of He in broad-lined SNe
is indeed in favor of standard stellar evolution models as 
it is difficult to remove all the helium layer from the progenitor
in many models \citep{Woosley1995,Yoon2010,Yoon2015}.
The observed diversity must reflect the variety
of the evolutionary paths, but the exact origin is not yet understood.
To understand the evolutionary paths to broad-lined SNe
and their variety, observations of more broad-lines SNe
from early phases will be important.

%Searching the companion star using late-phase observations
%will be encouraged when the SN component becomes sufficiently faint.
%Also, the environment of this SN must be of interest
%as metallicity is a key parameter for the progenitor scenario.

% \subsection{Explosion properties}

% Asymmetric signatures should be investigated for this object, using
% the late-time spectra, e.g., O~{\sc i}$\lambda\lambda$6300,6363
% line profile \citep{Maeda2008}.

%  On the other hand, the ejected $^{56}$Ni mass of SN 2016coi 
% was $\sim0.2M_{\odot}$, 
% which was smaller than those of other comparative SNe Ic-BL. These 
% facts may suggest that the new diversity was found for SNe Ic-BL.
 
%  We present the identification of SN Ib-BL for the first time.

%   \section{Conclusion}

 \acknowledgements
% We would like to thank Mohan Ganeshalingam for data of rise time versus 
% decline rate of light curves. 
% This research has been supported in part in part by Optical \& Near-infrared
% Astronomy Inter-University Cooperation Program (OISTER) and by the Grant-in-Aid for 
% Scientific Research from JSPS (23340048,24740117,23244030) and 
% MEXT (25103515,24103003). Y.M. and R.I. have been supported by the
% JSPS Research Fellowship for Young Scientists. 

%  \begin{ack}
  This work was supported
 by the Optical and Near-infrared Astronomy Inter-University Cooperation Program, 
 and the Hirao Taro Foundation of the Konan University Association 
 for Academic Research. 
 This work was partly supported by the Grant-in-Aid for 
 Scientific Research from JSPS (26800100, 15H02075, 15H00788). 
 The work by K.M. is partly supported by WPI Initiative, Mext, Japan.
% The work by K.M. is partly supported by World Premier 
% International Research Center Initiative (WPI Initiative), MEXT, Japan.
% Support for HK is provided by the Ministry of Economy, Development,
% and Tourism's Millennium Science Initiative through grant IC120009,
 %awarded to The Millennium Institute of Astrophysics, MAS. HK
 %acknowledges support by CONICYT through FONDECYT grant 3140563.
% Operation of ANIR on the miniTAO 1m telescope is supported by the
% Ministry of Education, Culture, Sports, Science, and Technology of
% Japan, Grant-in-Aid for Scientific Research (17104002, 20040003,
% 20041003, 21018003, 21018005, 21684006, 22253002, 22540258, and
% 23540261) from the Japan Society for the Promotion of Science (JSPS),
% and NAOJ Research Grant for Universities and Optical and Near-Infrared
% Astronomy Inter-University Cooperation Program.
%\end{ack}

% \bibliographystyle{apj}
% \bibliography{16coi,apj-jour}

\end{document}